\PassOptionsToPackage{table}{xcolor}
\pdfoutput=1
\documentclass[10pt, conference]{IEEEtran}
\usepackage{amsmath} 
\usepackage{amsthm} 
\usepackage{amssymb} 
\usepackage{graphicx} 
\usepackage{multicol} 
\usepackage{sparklines} 
\usepackage{subcaption}
\usepackage{cleveref}

\let\svthefootnote\thefootnote
\newcommand\blankfootnote[1]{%
  \let\thefootnote\relax\footnotetext{#1}%
  \let\thefootnote\svthefootnote%
}

\begin{document}

\title{How Research Software Engineers Can Support Scientific Software}
\author{\IEEEauthorblockN{Miranda Mundt, Evan Harvey}
\IEEEauthorblockA{mmundt@sandia.gov, eharvey@sandia.gov\\Department of Software Engineering and Research\\Sandia National Laboratories, 1611 Innovation Pkwy SE, Albuquerque, New Mexico 87123}}
\maketitle

\blankfootnote{Sandia National Laboratories is a multimission laboratory managed and operated by National Technology \& Engineering Solutions of Sandia, LLC, a wholly owned subsidiary of Honeywell International Inc., for the U.S. Department of Energy's National Nuclear Security Administration under contract DE-NA0003525. SAND2020-11113 C.
}

\section{Introduction}
We are research software engineers (RSE) and team members
in the Department of Software Engineering and Research at Sandia National
Laboratories, an organization which aims to advance software engineering in
the domain of computational science~\cite{deptwebsite}. Our team hopes to promote processes
and principles that lead to quality, rigor, correctness, and repeatability
in the implementation of algorithms and applications in scientific software
for high consequence applications.

Recently our department has undertaken the task of standardizing software
quality practices across the Center for Computing Research at Sandia with the
goal of using better tools and better practices to effect overall better
research. This is merely part of the effort. Because many
scientific software projects have limited funding, in lieu of having an RSE
readily available to assist with software engineering best practices (BPs),
we also strive to empower scientific researchers to properly leverage a
seemingly basic, albeit powerful, subset of software
tools and BPs in such a way that it promotes, rather than prohibits,
their research.

We use our experience to argue that there is a readily achievable
set of software tools and BPs with a large return on investment
that can be imparted upon scientific
researchers that will remarkably improve the quality of software and, as a
result, the quality of research.

\section{Scientific Software Development and RSEs}

D. F. Kelly once wrote, ``The engineering-software developer and software engineer's
lack of interest in each other's discipline is now firmly entrenched on both
sides of the chasm''~\cite{kelly2007software}. Kelly speaks of the vast chasm that separates the best
practices of software engineers from the domain-specific needs of researchers.
This gap stems primarily from a difference in priorities and values.

As Segal and Morris note, ``Most software developers have some idea of what a
human-resources or accounting package should do, and they feel they can
understand (perhaps with some effort) such packages' requirements''~\cite{segal2008developing}. This
differs significantly, however, from software generated as a result of scientific
research. The onset of many scientific software projects begins with a
previously unexplored research question, followed by exploratory attempts to
answer this research question, generally using software creation as a tool
to enable research.

Given the unique challenges present in scientific software development, many
common software engineering BPs cannot be applied without adaptation; that is,
common BPs can become stumbling blocks to ground-breaking research rather than facilitate it. For example, unit testing is cumbersome but can
significantly reduce debugging efforts. In scientific software, unit testing is
often only achievable by testing the end-result of top-level routines rather
than individually testing all internal and external routines.

As RSEs, we must connect researchers with the information and resources they
need to create software and enable scientific research, using language recognizable
and relatable to researchers to enable them to see the value
of the recommended tools and BPs. Many BPs may not fit the average scientific
software project without customization, but there is a subset of practices that
can benefit software development efforts of any size.

\section{Tools for Success}
Using the right tool often facilitates incorporating BPs. Below we suggest
tools with a large return on investment and their main selling points for
scientific software development in an effort to bring developer productivity
and software quality to scientific researchers.

\textbf{Version Control Systems}: In an anecdote by Gregory Wilson in American Scientist, he tells of introducing
version control systems (VCS) to a particle physicist who had written over 20,000
lines of code and who exclaimed, "Couldn't you have told me this three years
ago?"~\cite{wilson2006s}. As RSEs, the use of VCS is as natural as breathing,
but this may not be the case for the average researcher. We surmise that
for researchers, the three main BPs facilitated by VCS adoption are
data protection, collaboration, and scalability as the project matures.

The use of version control tools enables the BP of data protection by
encouraging a developer workflow that results in frequent backups and guards
against hardware and human errors. Additionally, VCS naturally fosters collaboration,
particularly when it comes to larger team sizes. An added benefit appears once a
customer becomes involved as many VCS support requirements and issue trackers to
manage customer requests and concerns. Finally, by scalability, we mean that version
control tools can frequently grow to meet project needs.
Since researchers will be accustomed to completing work using version control
tools from day one, it will be less of a lift to implement more rigorously defined
workflow standards (such as use of branches, forks, merge requests, etc.) as the
project matures.

A barrier to entry for VCS exists due to lack of formalized training. Many
practitioners must learn on the job. In our department, we created a formalized
VCS training program, given over a series of weeks, in an effort to empower scientific
software developers with VCS BPs. If resources are not available to create
such a program, RSEs can always point researchers to established training
material, such as through the Best Practices for HPC Software Developers webinar
series~\cite{ecptraining} or Better Scientific Software~\cite{bsswexample}.

\textbf{Testing}: There is no question about it - testing in scientific software is challenging.
Many times, the ``correct'' answer to a research problem is already an estimation
based on a previous algorithm or is altogether completely unknown. A standard BP for any
software project is the assertion for and reliance on testing. Continuous
integration (CI) tools lay a strong foundation for building reliable, and
subsequently reproducible, software where productivity is fostered.

In a recent whitepaper by fellow RSEs Aaron Levine and Jim Willenbring, the
authors describe development challenges for two scientific software projects.
Both software projects suffered from a lack of software stability, and
"[f]requently developers would unwittingly push broken code into the repository
that caused nightly tests to fail. Recovery from this broken state usually took
upwards of 3 days to correct"~\cite{levine2020rapids}. The authors then describe how
development challenges were addressed with a CI tool to the ultimate
betterment of the projects and their results.

For any software project, scientific or otherwise, a main goal is to promote
reliability and reproducibility, that is, the same methods and input data
result in the same computed values~\cite{krafczyk2019scientific}. Reproducibility in scientific software
discoveries is the penultimate achievement of any scientific research project.
A main argument for scientific researchers to employ CI tools is that they improve
software reliability in order to achieve reproducibility.
They do so by automatically enforcing that tests
pass before code changes are introduced into the software, thus resulting in a
more consistently functioning software that produces consistent output.

Another main selling point is that CI improves developer productivity by
automating validation and functionality testing
and minimizing the amount of time the software spends in a broken state. The
more a software works, the more developers and users will trust and continue to
utilize the software. This fosters better collaboration and working
relationships between developers and users.

It is true that CI tools have a non-trivial upfront cost. Particularly, the
implementation of CI tools requires someone with the knowledge of their use and
maintenance. Setting aside a small amount of money to fund an RSE to set up the
CI and coach the scientific researcher on its use would be our recommended
strategy to overcome this barrier. Our department at Sandia, as an example, is
able to ``loan'' an RSE either for a short or long period of time to a project,
dependent on funding availability. Our experience through this model is that,
though time and resources are always limited, the investment in CI often results in improved
developer happiness, better collaboration, and more time to focus on research.

\section{Conclusion}
When it comes to scientific software, researchers face unique challenges, from
lack of strictly defined requirements to lack of funding. As members of the
Department of Software Engineering and Research, we aim to advance software engineering in
the domain of computational science. In this paper, we argued that setting researchers
on the path to success using easy to adopt and large return on investment tools,
such as VCS and CI, shrinks the gap between scientific researchers and RSEs.
This effort builds trust and understanding between the two communities while enabling
scientific researchers to use better tools and better practices to effect overall better
research. It is up to us to present these tools to the researchers who need
them and integrate our values for the betterment of software and scientific research.

\bibliographystyle{IEEEtran}
\bibliography{rsehpc}

\end{document}